\documentclass[graybox]{svmult}

\usepackage{mathptmx}       
\usepackage{helvet}         
\usepackage{courier}        
\usepackage{type1cm}        
\usepackage{makeidx}         
\usepackage{graphicx}        
\usepackage{multicol}        
\usepackage[bottom]{footmisc}

\makeindex             

\begin{document}

\title*{Comparing Various Approaches to Simulating the Formation of Shell Galaxies}
\titlerunning{New Approaches to Simulating the Formation of Shell Galaxies}
\author{Ivana Ebrov\'{a}, Kate\v{r}ina Barto\v{s}kov\'{a}, Bruno Jungwiert, Lucie J\'{i}lkov\'{a}, Miroslav K\v{r}\'{i}\v{z}ek}
\authorrunning{Ebrov\'{a} et al.}
\institute{Ivana Ebrov\'{a}, Bruno Jungwiert, and Miroslav K\v{r}\'{i}\v{z}ek 
\at Astronomical Institute, Academy of Sciences of the Czech Republic; Faculty of Mathematics and Physics, Charles University in Prague, \email{ivana@ig.cas.cz}
\and Kate\v{r}ina Barto\v{s}kov\'{a}
\at Faculty of Science, Masaryk University, Brno, Czech Republic; Astronomical Institute, Academy of Sciences of the Czech Republic
\and Lucie J\'{i}lkov\'{a} 
\at ESO Santiago, Chile; Faculty of Science, Masaryk University, Brno, Czech Republic
}

\maketitle

\vskip-1.2truein

\abstract{The model of a radial minor merger proposed by~\cite{quinn84}, which successfully reproduces the observed regular shell systems in shell galaxies, is ideal for a test-particle simulation. We compare such a simulation with a self-consistent one. They agree very well in positions of the first generation of shells but potentially important effects -- dynamical friction and gradual decay of the dwarf galaxy -- are not present in the test-particle model, therefore we look for a proper way to include them.}

\section*{Tides and dynamical friction in  test particle simulations}
\label{sec1:intro}
We model the luminous and dark matter components of a host giant elliptical (gE) and a dwarf elliptical galaxy by analytical potentials. In the simplest model, we assume the dwarf galaxy, filled with millions of test particles, to be ripped apart instantly when it comes close to the center of the gE galaxy. Its stars begin to oscillate in the potential of the host galaxy and produce shells at their turning points.

Such a setup allows us to use large numbers of particles and so to gain sufficient contrast to detect all the shells in the simulation, also to investigate the kinematic footprint in spectral lines (see~\cite{huevo} and~\cite{lucka}) and explore a large parameter space. This would be very time consuming for large sets of self-consistent simulations.

\begin{figure}[t]
%
\begin{center}
\includegraphics[scale=.49]{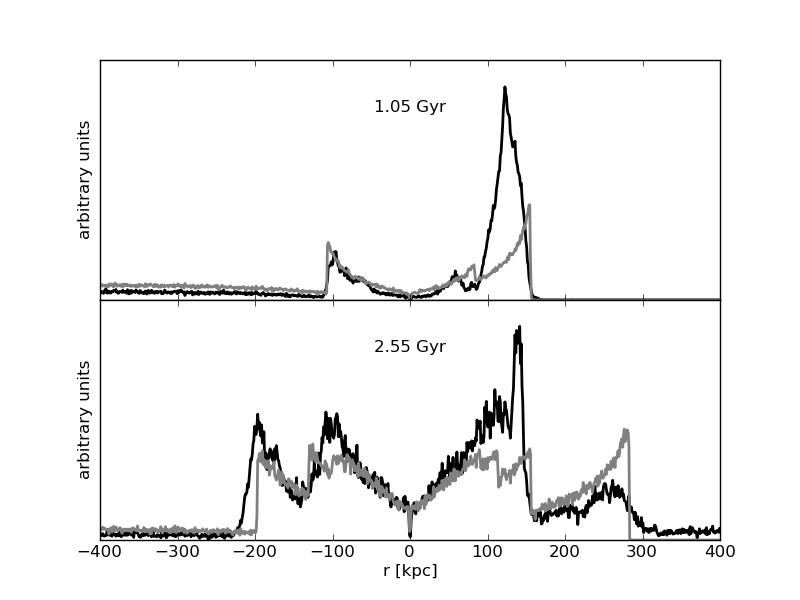}
\end{center}
\caption{The graph shows the comparison of histograms of radial distances of shells' particles (centered on the host galaxy)  in the self-consistent (black) and test-particle (grey) simulations at two different time steps. Time equal zero corresponds to the passage of the dwarf galaxy through the center of the host galaxy. Both simulations have the same initial conditions. Notice that the brightness of the outmermost shell (at 280 kpc at 2.55 Gyr) is supressed in the self-consistent simulation. This effect has been successfully simulated in the improved test-particle simulations; see \cite{usa}.}
\label{fig:ebrova_1}       
\end{figure}

Surprisingly, the agreement with a self-consistent simulation (for more details see~\cite{katka}) turns out to be very good especially in the positions of shells (see Fig.~\ref{fig:ebrova_1} and~\cite{katka}). But the simple model does not involve effects like dynamical friction and gradual decay of the dwarf galaxy, so that it cannot simulate phenomena seen in self-consistent simulations: the next generation of shells (see~\cite{katka}) and lowering of brightness of shells. We thus look for a middle way, where we can still have the large contrast available through the use of test particles, yet include some of the more complicated effects to make it more realistic.   

In this improved test-particle simulation we use our version of enhanced Chandrasekhar formula with variable Coulomb logarithm to include dynamical friction, and we also introduced a gradual decline of the mass parameter of the dwarf galaxy potential to better imitate the evolution of tides. For details, see~\cite{usa} and \cite{ivana11}.

\begin{acknowledgement}
We acknowledge the support by the grant No. 205/08/H005 (Czech Science Foundation), the project SVV 261301 (Charles University 
in Prague), EAS grant covering  registration  fee at JENAM 2010,  the grant MUNI/A/0968/2009  (Masaryk University  in Brno),  the 
grant LC06014 'Center for Theoretical Astrophysics' (Czech Ministry of Education) and research plan AV0Z10030501 (Academy of 
Sciences of the Czech Republic).
\end{acknowledgement}


\end{document}